\documentclass[pdflatex,sn-mathphys-num]{sn-jnl}

\usepackage{graphicx}%
\usepackage{multirow}%
\usepackage{amsmath,amssymb,amsfonts}%
\usepackage{amsthm}%
\usepackage{mathrsfs}%
\usepackage[title]{appendix}%
\usepackage{xcolor}%
\usepackage{textcomp}%
\usepackage{manyfoot}%
\usepackage{booktabs}%
\usepackage{algorithm}%
\usepackage{algorithmicx}%
\usepackage{algpseudocode}%
\usepackage{listings}%
\usepackage{booktabs}
\usepackage{hyperref}
\usepackage{pifont}

\newcommand{\xmark}{\ding{55}}

\theoremstyle{thmstyleone}%
\theoremstyle{thmstyletwo}%

\theoremstyle{thmstylethree}%

\begin{document}

\title[Article Title]{ST-DAI: Single-shot 2.5D Spatial Transcriptomics with Intra-Sample Domain Adaptive Imputation for Cost-efficient 3D Reconstruction}


\author*[1,4]{\fnm{Jiahe} \sur{Qian}}\email{jiahe.qian@northwestern.edu}

\author[1]{\fnm{Yaoyu} \sur{Fang}}\email{yaoyu.fang@northwestern.edu}

\author[2]{\fnm{Xinkun} \sur{Wang}}\email{xinkun.wang@gmail.com}

\author[3]{\fnm{Lee A.} \sur{Cooper}}\email{lee.cooper@northwestern.edu}

\author[1]{\fnm{Bo} \sur{Zhou}}\email{bo.zhou@northwestern.edu}

\affil[1]{\orgdiv{Department of Radiology}, \orgname{Northwestern University}, \orgaddress{\street{Chicago}, \postcode{60611}, \state{IL}, \country{USA}}}

\affil[2]{\orgdiv{Department of Cell and Developmental Biology}, \orgname{Northwestern University}, \orgaddress{\street{Chicago}, \postcode{60611}, \state{IL}, \country{USA}}}

\affil[3]{\orgdiv{Department of Pathology}, \orgname{Northwestern University}, \orgaddress{\street{Chicago}, \postcode{60611}, \state{IL}, \country{USA}}}

\affil[4]{\orgdiv{Institute of Automation},
          \orgname{Chinese Academy of Sciences},
          \orgaddress{\street{Beijing}, \postcode{100190}, \country{China}}}


\abstract{For 3D spatial transcriptomics (ST), the high per-section acquisition cost of fully sampling every tissue section remains a significant challenge.  Although recent approaches predict gene expression from histology images, these methods require large external datasets, which leads to high-cost and suffers from substantial domain discrepancies that lead to poor generalization on new samples.  In this work, we introduce ST-DAI, a single-shot framework for 3D ST that couples a cost-efficient 2.5D sampling scheme with an intra-sample domain-adaptive imputation framework.  First, in the cost-efficient 2.5D sampling stage, one reference section (central section) is fully sampled while other sections (adjacent sections) is sparsely sampled, thereby capturing volumetric context at significantly reduced experimental cost.  Second, we propose a single-shot 3D imputation learning method that allows us to generate fully sampled 3D ST from this cost-efficient 2.5D ST scheme, using only sample-specific training. We observe position misalignment and domain discrepancy between sections. To address those issues, we adopt a pipeline that first aligns the central section to the adjacent section, thereafter generates dense pseudo-supervision on the central section, and then performs Fast Multi-Domain Refinement (FMDR), which adapts the network to the domain of the adjacent section while fine-tuning only a few parameters through the use of Parameter-Efficient Domain-Alignment Layers (PDLs). During this refinement, a Confidence Score Generator (CSG) reweights the pseudo-labels according to their estimated reliability, thereby directing imputation toward trustworthy regions. Our experimental results demonstrate that ST-DAI achieves gene expression prediction performance comparable to fully sampled approaches while substantially reducing the measurement burden.}




\maketitle

\section{Introduction}
\begin{figure*}
    \centering
    \includegraphics[width=1\linewidth]{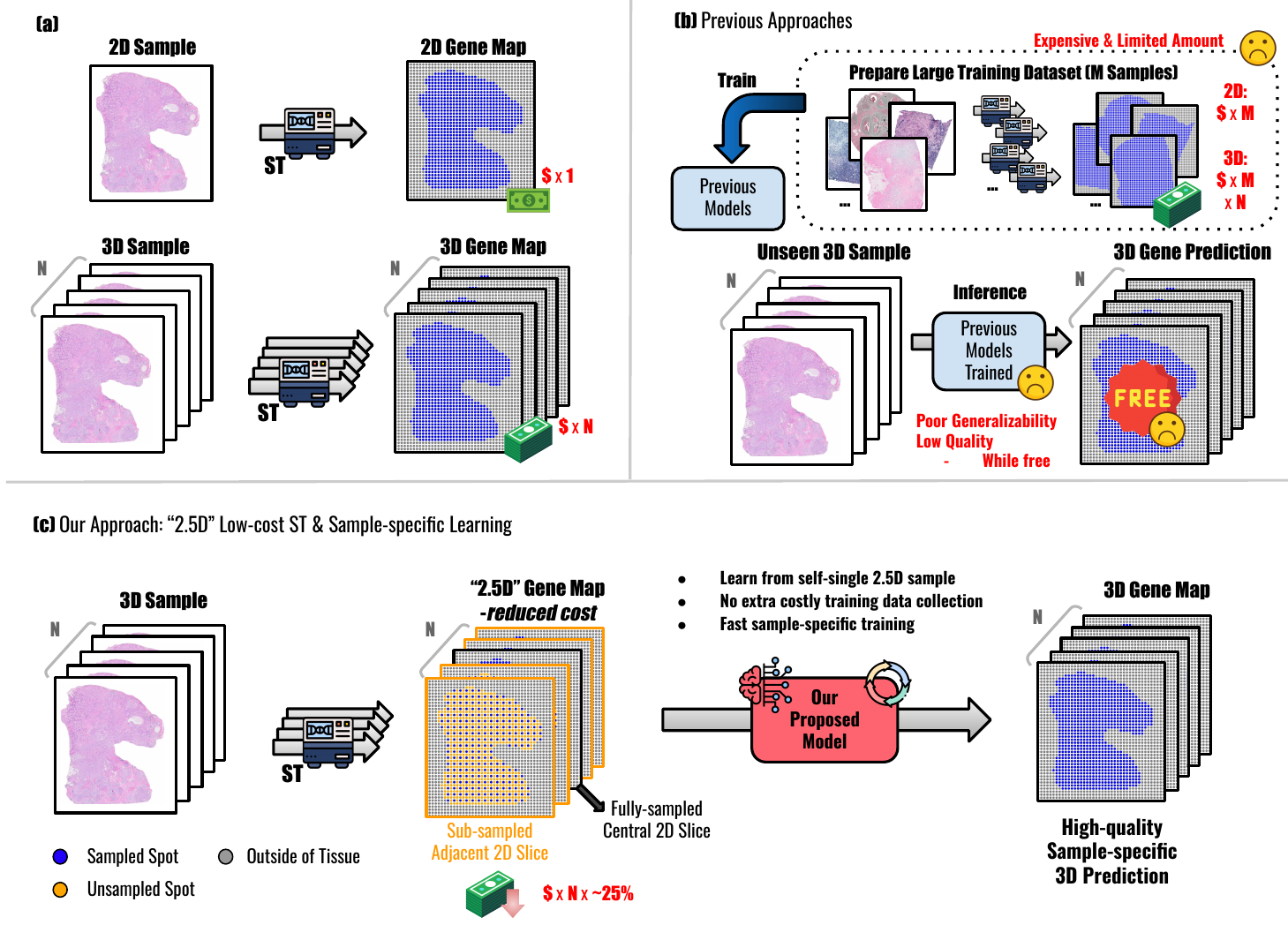}
    \caption{Illustration of three distinct sampling settings for 3D ST and its prediction. The first setting involves full sampling of every section, yielding high accuracy at the expense of prohibitively high sampling costs. The second setting relies on training deep networks with large external datasets to predict gene expression from histology images, thereby eliminating within-sample sampling but incurring high aggregate sampling costs and suffering from poor generalization due to domain shifts. In contrast, our proposed setting (ST-DAI) requires full sampling of only a central section and sparse sampling of adjacent sections, substantially reducing sampling costs while maintaining high prediction accuracy and obviating the need for external training data.}
    \label{fig:Problem1}
\end{figure*}

Mapping gene expression in three dimensions (3D) within intact tissues is fundamental for understanding spatial biology and disease, yet obtaining such data at scale remains challenging. Recent Spatial Transcriptomics (ST) technologies make it possible to measure mRNA expression while preserving spatial context in tissue sections \cite{Marx2021, Stahl2016}. However, these techniques are typically applied on individual two-dimensional (2D) sections, and extending them to whole 3D samples naively would require profiling every serial section of a tissue, as shown in Figure~\ref{fig:Problem1}(a). This brute-force approach of full sampling across all sections can indeed produce an accurate 3D gene expression atlas \cite{Rodriques2019}, but it is prohibitively expensive and labor-intensive. Even with high-throughput methods like high-definition ST and Slide-seq \cite{Wang2018, Rodriques2019}, the cost and effort scale linearly with the number of sections, making comprehensive 3D profiling impractical for large organs, tissues, or clinical cohorts. In summary, exhaustive within-sample sequencing is effective but not scalable.

Previous efforts have attempted to mitigate the sequencing burden by learning computational surrogates of gene expression, yet two principal challenges remain.  First, existing methods train deep neural networks on massive external collections of image–transcriptome pairs \cite{Schmauch2020, Xie2023, Coudray2018, Campanella2019, zhu2024asign, yang2023exemplar, ortiz2020molecular, andersson2021spatial, wu2021single}.  Constructing such corpora entails pairing stained tissue slides with bulk or spatial RNA-sequencing measurements drawn from numerous specimens \cite{Stahl2016, Rodriques2019}, which imposes substantial sequencing and annotation costs that merely redistribute rather than eliminate experimental expenditure \cite{Tellez2019}.  Second, although a model trained on these external data can be applied directly to 3D samples, its prediction accuracy on unseen tissues often degrades significantly due to domain changes induced by variations in processing protocols, staining conditions, and intrinsic biological heterogeneity \cite{Saillard2023, Stacke2021}. These constraints underscore the importance of in-distribution learning, wherein the model is calibrated solely on data acquired from the specimen of interest, i.e., single-shot or sample-specific, thereby eliminating dependence on large external datasets and intrinsically reducing cross-domain generalisation error.

\begin{figure}[htbp]
    \centering
    \includegraphics[width=0.8\linewidth]{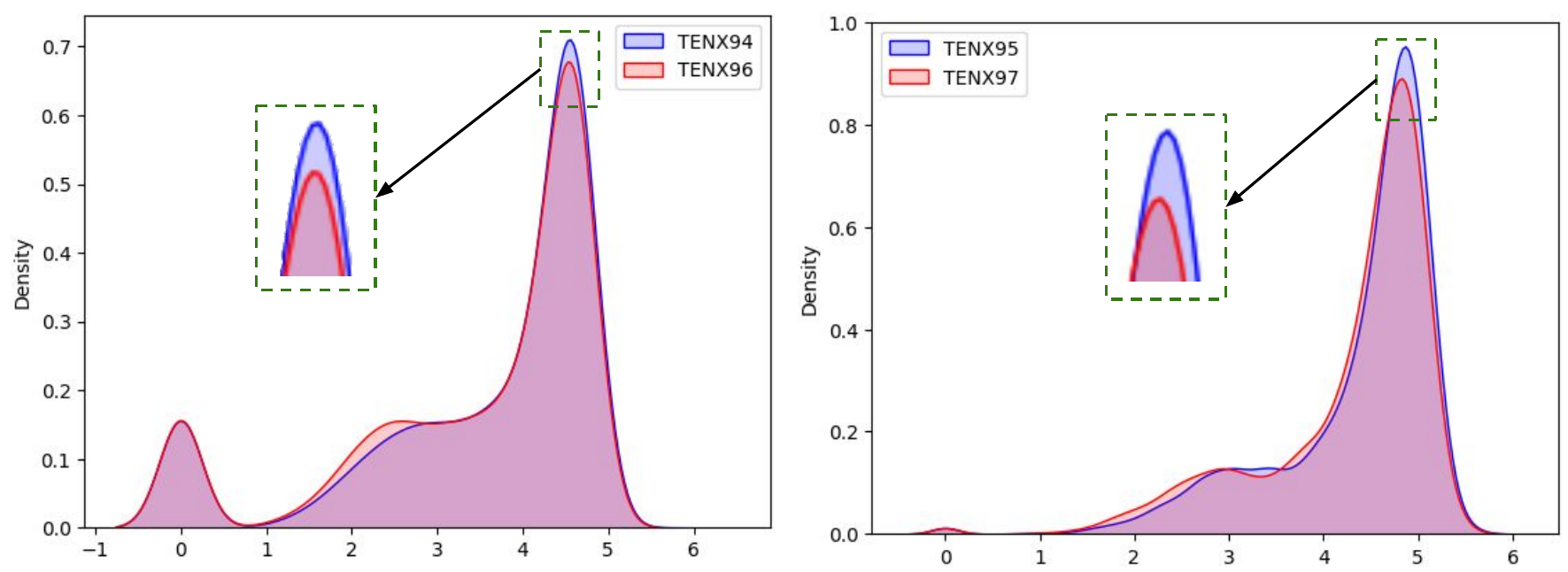}
    \caption{Probability density functions of ERBB2 expression for two pairs of spatially adjacent breast cancer sections with domain gap observed. The left pair comprises TENX95 and TENX97, whereas the right pair comprises TENX94 and TENX96. Each pair is derived from a distinct patient with invasive ductal carcinoma. The horizontal axis represents the quantified ERBB2 expression level.}
    \label{fig:Problem2}
\end{figure}

While the sample-specific single-shot setting can significantly reduce the cross-sample domain gap, there remain several challenges in multi-section 3D ST settings. First, there exists an intra-sample domain gap between consecutive sections, manifested as both positional misalignment and shifts in expression distributions. Mechanical deformation during sectioning, variations in section orientation, and intrinsic biological heterogeneity cause neighbouring sections to diverge, so that each section effectively constitutes a slightly shifted domain.  Figure~\ref{fig:Problem2} illustrates an example of this gap by comparing the probability density functions of ERBB2 expression in two pairs of consecutive breast cancer sections, namely TENX95 versus TENX97 and TENX94 versus TENX96, where each pair is obtained from two different patients with invasive ductal carcinoma. The differences between the distributions highlight the necessity of section-level domain adaptation. Second, most existing frameworks assume that no expression information is available on adjacent sections, forcing the model to depend exclusively on the fully sequenced central section.  Introducing even sparse measurements on neighbouring sections would provide valuable supervision and could potentially improve the fidelity of 3D reconstruction.

To overcome these limitations, we introduce \textbf{ST-DAI}, a sample-specific single-shot solution that unifies a cost-efficient \(2.5\mathrm{D}\) sampling protocol with a domain-adaptive framework for 3D intra-sample imputation, as summarised in Figure~\ref{fig:Problem1}(c). ST-DAI combines two complementary advances:

\noindent\textbf{(1) A cost-efficient 2.5D sampling scheme fully profiles one representative tissue section while sparsely sampling its immediate neighbours.}\quad
The central section is sequenced at high spatial resolution so that the complete transcriptomics is captured at every location, whereas a predefined grid on the adjacent sections collects measurements from only a subset of sites.  The proposed sampling protocol markedly lowers sequencing expenditure while still producing one fully profiled central section that supplies comprehensive transcriptomic detail.  Combining this rich reference with the sparsely measured adjacent sections permits sample-specific reconstruction of the entire 3D gene-expression landscape.

\noindent\textbf{(2) A coupled intra-sample domain-adaptive imputation framework transforms the heterogeneous \(2.5\mathrm{D}\) observations into a coherent three-dimensional expression volume.}\quad
The process starts with Cross-section Alignment (CSA), which registers the fully sampled central section to every sparsely sampled adjacent section, thereby unifying all data in a common spatial frame.  A Pseudo Map Network pretrained on the central section then propagates transcriptomic information by generating pseudo predictions for the unsampled regions of each adjacent section.  Next, Fast Multi-Domain Refinement (FMDR) adapts the model to section-specific characteristics by updating only the Parameter-Efficient Domain-Alignment Layers together with the terminal layer, while the shared backbone remains frozen.  During 3D inference, a Data Consistency Operation (DCO) fuses these refined predictions with the experimentally observed measurements, producing an internally consistent and high-fidelity volumetric gene-expression reconstruction.

Comprehensive experiments on real ST sample benchmarks demonstrate that ST-DAI attains high reconstruction fidelity while requiring only a fraction of the sequencing effort, underscoring its potential practicality and scalability for volumetric ST studies.

\section{Methods}
ST-DAI comprises two tightly coupled components. The first is a cost-efficient 2.5D sampling strategy that acquires a complete ST profile on a single central section and sparse measurements on every other adjacent section, thereby lowering sequencing demand while maintaining volumetric coverage.  The second is a intra-sample domain-adaptive imputation methodology that converts these 2.5D heterogeneous observations into a coherent 3D gene-expression reconstruction.

The remainder of this section is organized as follows.  Section~\ref{method:sampling} details the 2.5D sampling strategy.  Section~\ref{method:overview} outlines the overall intra-sample domain-adaptive imputation framework. Sections~\ref{method:csg} and~\ref{method:al} describe two key algorithmic modules: the Confidence Score Generator (CSG), which estimates the reliability of pseudo labels, and the Parameter-Efficient Domain-Alignment Layer (PDL), which aligns feature distributions across sections with minimal trainable parameters.  Section~\ref{method:dco} defines the Data Consistency Operation (DCO) that fuses predictions with observed measurements during inference.  Section~\ref{method:loss} specifies the loss functions used for optimization. Section~\ref{method:implementation} summarises network architecture and implementation details, and Section~\ref{method:datasets} introduces the datasets and evaluation metrics employed in our experiments.

\begin{figure*}
    \centering
    \includegraphics[width=1\linewidth]{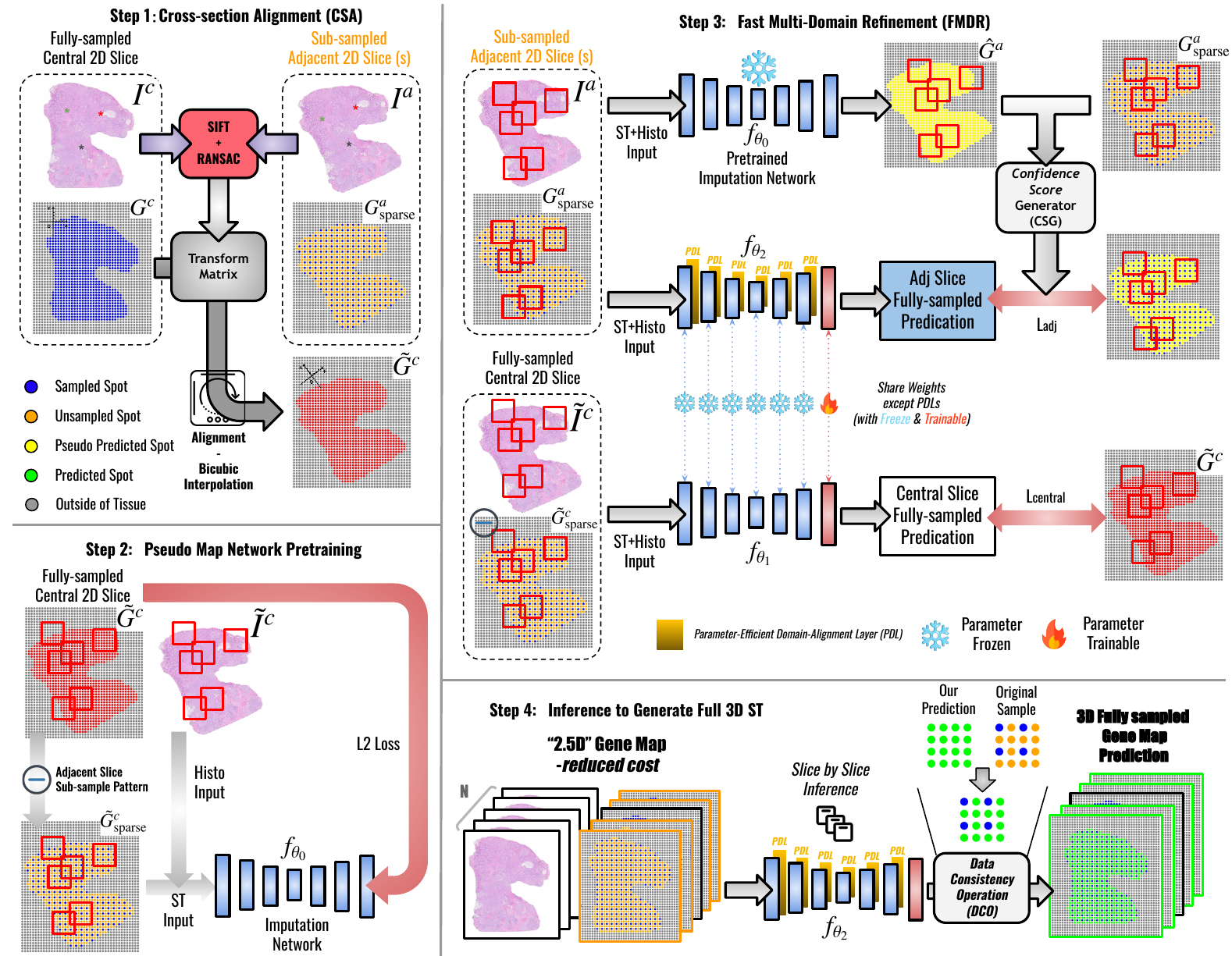}
    \caption{Overview of the 3D imputation framework of ST-DAI. The method consists of four stages: (1) Cross-section Alignment (CSA) establishes a common spatial reference between the fully sampled central section and the sparsely sampled adjacent sections, (2) Pseudo Map Network Pretraining leverages the central section to generate pseudo maps and for pre-training a imputation network, (3) Fast Multi-Domain Refinement (FMDR) employs a dual-branch strategy enhanced with Parameter-Efficient Domain-Alignment Layers (PDL) and a Confidence Score Generator (CSG) to enable domain-adaptive fine-tuning while limiting training to the final layer, (4) Final Inference applies a Data Consistency Operation (DCO) to integrate observed and predicted data for full 3D imputation from 2.5D sample.}
    \label{fig:Method}
\end{figure*}

\subsection{Cost-efficient 2.5D Sampling}

\label{method:sampling}

Consider an ordered set of equidistant histological sections \(\{S_{1},S_{2},\dots,S_{n}\}\) extracted from a 3D tissue block.  One index \(k\in\{1,\dots,n\}\) is designated such that \(S_{k}\) serves as the central section. ST sequencing is performed exhaustively on \(S_{k}\), yielding a complete expression measurement at every spatial location \(\mathbf{x}\in\Omega\subset\mathbb{R}^{2}\).

For each remaining section \(S_{j}\,(j\neq k)\), gene expression is acquired only on a regular sampling grid
\[
\mathcal{G}_{j}=\bigl\{\mathbf{x}_{p}\bigr\}_{p=1}^{K}\subset\Omega,
\]
where the grid spacing is fixed at two pixels in both axial directions, so exactly one position is selected within every \(2\times2\) block of the image lattice. The resulting set of expression measurements \(\{\,y_{j}(\mathbf{x}_{p})\mid\mathbf{x}_{p}\in\mathcal{G}_{j}\,\}\) provides a sparse yet informative representation of each adjacent section.  Because all sections except \(S_{k}\) follow the same grid pattern, the sampling density is spatially homogeneous, facilitating subsequent cross-section alignment and joint modelling.

By fully profiling a single section and sparsely sampling all others, the proposed strategy substantially curtails sequencing cost while retaining sufficient spatial coverage to support accurate 3D reconstruction.

\subsection{2.5D Intra-Sample Domain-Adaptive Imputation}
\label{method:overview}

As shown in \ref{fig:Method}, our 3D imputation framework for ST-DAI comprises four sequential stages: Cross-section Alignment (CSA), Pseudo Map Network Pretraining, Fast Multi-Domain Refinement (FMDR), and 3D Gene Map Prediction.  
For notation, let \(I^{c}\in\mathbb{R}^{H\times W\times 3}\) and \(G^{c}\in\mathbb{R}^{H\times W}\) denote, respectively, the histology image and the fully observed gene–expression map of the central section.  
For each adjacent section, let \(I^{a}\in\mathbb{R}^{H\times W\times 3}\) be its histology image and let \(G^{a}_{\text{sparse}}\) represent the corresponding sparsely measured expression values.

\noindent\textbf{Cross-section Alignment (CSA).}\quad
Robust feature-based registration is performed by extracting Scale-Invariant Feature Transform (SIFT) keypoints and estimating the geometric transformation with the Random Sample Consensus (RANSAC) algorithm, thereby aligning the fully sampled central section to each sparsely sampled neighbouring section \cite{Lowe2004, Fischler1981}.  After applying the estimated transformation, the warped central-section image and its gene-expression map are denoted by \(\tilde{I}^{c}\) and \(\tilde{G}^{c}\), respectively.

\noindent\textbf{Pseudo Map Network Pretraining.}\quad
A pseudo-map network \(f_{\theta_{0}}\) is first trained on the masked central-section expression map \(\tilde{G}^{c}_{\text{sparse}}\), which is obtained by sampling \(\tilde{G}^{c}\) on exactly the same sparse grid applied to the adjacent sections. The trained network then generates pseudo labels \(\hat{G}^{a}=f_{\theta_{0}}\bigl(I^{a}, G^{a}_{\text{sparse}}\bigr)\) for the unmeasured regions of every adjacent section \(a\).

\noindent\textbf{Fast Multi-Domain Refinement (FMDR).}\quad
To mitigate inter-section domain discrepancies while maintaining computational efficiency, a dual-branch architecture is adopted. The central branch, denoted \(f_{\theta_{1}}\), processes \(\bigl(\tilde{I}^{c}, \tilde{G}^{c}_{\text{sparse}}\bigr)\) and keeps the backbone entirely frozen, updating solely its terminal layer. The adjacent-section branch, denoted \(f_{\theta_{2}}\), processes \(\bigl(I^{a}, G^{a}_{\text{sparse}}\bigr)\) and is fine-tuned only in its Parameter-Efficient Domain-Alignment Layers (PDL) together with its own terminal layer, thereby recalibrating feature statistics with minimal additional parameters. A Confidence Score Generator (CSG) assigns reliability weights to pseudo labels on a pixel basis, guiding the optimisation towards trustworthy regions. This design realises rapid convergence and robust domain adaptation with minimal learnable parameters.

\noindent\textbf{3D Gene Map Prediction.}\quad
After refinement, the model outputs dense expression estimates \(G^{a}_{\text{pred}}\) for the adjacent section \(a\). Finally, a Data Consistency Operation (DCO) then replaces each predicted value at a measured location with its corresponding observation, yielding section-level results \(G^{a}_{\text{final}}\) and producing the final 3D gene-expression tensor \(G_{\text{final}}\).

Detailed descriptions of PDL, CSG, DCO, loss functions, implementation specifics, and dataset and evaluation metrics are provided in Sections \ref{method:al}, \ref{method:csg}, \ref{method:dco}, \ref{method:loss}, \ref{method:implementation}, and \ref{method:datasets}.

\subsection{Confidence Score Generator}\label{method:csg}

Confidence Score Generator (CSG) assigns a reliability weight to every pixel in an adjacent section by first evaluating the observed locations and then propagating these assessments to the entire image.  
For each section, let the binary mask 
\[
M_i^{\text{vis}} =
\begin{cases}
1, & \text{if pixel } i \text{ is directly observed},\\[2mm]
0, & \text{otherwise},
\end{cases}
\]
indicate whether pixel \(i\) belongs to the observed set \(\mathcal{S}\).  
For every observed pixel \(j\in\mathcal{S}\), the prediction error is computed as the Euclidean distance between the pseudo prediction \(\hat{G}^{a}_{j}\) and the measured gene-expression value \(G^{a}_{j}\),
\[
E_{j}= \lVert \hat{G}^{a}_{j}-G^{a}_{j}\rVert_{2},
\]
and \(\max_{k\in\mathcal{S}}E_{k}\) denotes the maximum error within the observed set.  
An initial confidence score for each observed pixel is defined as
\[
w_{j}^{\text{obs}} = 1-\frac{E_{j}}{\max_{k\in\mathcal{S}}E_{k}}, \qquad j\in\mathcal{S},
\]
so that smaller errors correspond to higher confidence.  
These scores are then extended to all \(N\) pixels by bicubic interpolation,
\[
w_{i}= \mathcal{I}\!\bigl(\{w_{j}^{\text{obs}}\}_{j\in\mathcal{S}}\bigr), \qquad i=1,\dots ,N,
\]
where \(\mathcal{I}(\cdot)\) denotes the interpolation operator.  

To preserve the authority of experimentally sampled sites, we first impose a hard constraint that assigns unit confidence to every directly measured pixel,
\[
w_{i} =
\begin{cases}
1, & i \in \mathcal{S},\\[2mm]
w_{i}, & i \notin \mathcal{S},
\end{cases}
\]
thereby anchoring the weighting scheme to the most reliable observations.  
Next, to maintain numerical stability, the entire confidence map is renormalised to unit mean,
\[
\tilde{w}_{i}= \frac{w_{i}}{\bar{w}},\qquad \bar{w}= \frac{1}{N}\sum_{k=1}^{N} w_{k},
\]
 
The resulting pixel-wise confidence map is applied multiplicatively to the reconstruction loss, so locations that carry reliable supervision contribute more strongly to the optimisation process. Consequently, the network is guided away from uncertain pseudo labels and toward improved accuracy in high-resolution gene-expression prediction.

\subsection{Parameter-Efficient Domain-Alignment Layer}
\label{method:al}

The Parameter-Efficient Domain-Alignment Layer (PDL) aims to facilitate rapid domain adaptation of the adjacent section branch by introducing only a negligible number of additional parameters. PDLs are inserted after any intermediate operation within the pretrained backbone, thereby preserving the existing feature distribution while providing local domain‑specific calibration.

Let \(\mathbf{f}\in\mathbb{R}^{C\times H\times W}\) denote an intermediate feature tensor and let \(\mathbf{a},\mathbf{b}\in\mathbb{R}^{C}\) be trainable channel‑wise vectors. PDL applies the affine transformation
\begin{equation}
\hat{\mathbf{f}} \;=\; ( \mathbf{1} + \mathbf{a}) \odot \mathbf{f} \;+\; \mathbf{b},
\label{eq:pdl_vector}
\end{equation}
where \(\odot\) indicates channel‑wise multiplication. Component‑wise this becomes
\begin{equation}
\hat{f}_{c,h,w} \;=\; \bigl(1 + a_{c}\bigr)\,f_{c,h,w} \;+\; b_{c},
\label{eq:pdl_scalar}
\end{equation}
with \(c\in\{1,\dots,C\}\) and spatial index \((h,w)\). The vectors \(\mathbf{a}\) and \(\mathbf{b}\) are initialized to zero, guaranteeing that the transformation is identity at the onset of fine‑tuning. During adaptation, only \(\mathbf{a}\), \(\mathbf{b}\) and the network’s terminal layer are updated; all preceding parameters remain frozen. Consequently, the learnable parameter count introduced by PDL is insignificant relative to the frozen portion of the model, yet it affords sufficient flexibility for subtal and precise domain alignment. By limiting optimization to this small set of parameters, PDL significantly accelerates training while achieving rapid domain alignment.

\subsection{Data Consistency Operation}
\label{method:dco}

The Data Consistency Operation (DCO) enforces that experimentally measured values are embedded directly into the final 3D gene-expression volume.  
For each pixel \(i\) in an adjacent section \(a\), let \(G^{a}_{i}\) denote the measured expression value when pixel \(i\) is sampled, and let \(G^{a}_{\text{pred},\,i}\) denote the network prediction.  DCO is defined as
\begin{equation}
G^{a}_{\mathrm{final}}(i)=
\begin{cases}
G^{a}_{i}, & \text{if pixel } i \text{ is sampled},\\[5mm]
G^{a}_{\text{pred},\,i}, & \text{otherwise}.
\end{cases}
\end{equation}

This rule guarantees that the final output \(G^{a}_{\mathrm{final}}\) faithfully preserves all real measurements, while filling unsampled regions with the network’s predictions.

\subsection{Objective Functions}
\label{method:loss}

Throughout both the Pseudo-Map Network pretraining and the subsequent FMDR stage, supervision for the fully sampled central section is provided using a standard L2 loss. Let \(G^{c}_{i}\) denote the measured gene-expression value at pixel \(i\) in the central section, and let  
\(\bigl[f_{\theta_{0}}(I^{c},G^{c}_{\text{sparse}})\bigr]_{i}\) and  
\(\bigl[f_{\theta_{1}}(I^{c},G^{c}_{\text{sparse}})\bigr]_{i}\)  
denote the corresponding predictions before and after refinement, respectively. The supervised losses for the central section in the two stages are
\begin{align}
L_{\text{central}}^{\text{pseudo}} &= \frac{1}{N_c}\sum_{i=1}^{N_c}\bigl\|G^{c}_{i}-\bigl[f_{\theta_{0}}(I^{c},G^{c}_{\text{sparse}})\bigr]_{i}\bigr\|_{2}^{2},\label{eq:loss-pre}\\
L_{\text{central}} &= \frac{1}{N_c}\sum_{i=1}^{N_c}\bigl\|G^{c}_{i}-\bigl[f_{\theta_{1}}(I^{c},G^{c}_{\text{sparse}})\bigr]_{i}\bigr\|_{2}^{2},\label{eq:loss-fmdr}
\end{align}
where \(N_c\) is the total number of pixels in the central section.

For an adjacent section, the Pseudo Map Network produces pseudo labels \(\hat{G}^{a}_{i}\), and the Confidence Score Generator supplies a normalised confidence weight \(\tilde{w}_{i}\) for each pixel.  
The loss on the adjacent section is then computed as a weighted L2 loss:
\begin{equation}
L_{\text{adj}} = \frac{1}{N_a} \sum_{i=1}^{N_a} \tilde{w}_{i}\, \bigl\| \hat{G}^{a}_{i} - \bigl[f_{\theta_{2}}(I^{a},G^{a}_{\text{sparse}})\bigr]_{i} \bigr\|_{2}^{2},
\end{equation}
where \(N_a\) is the total number of pixels in the adjacent section.

The overall objective during the FMDR is defined as the sum of these two loss terms:
\begin{equation}
\mathcal{L} = L_{\text{central}} + L_{\text{adj}}.
\end{equation}
Because the confidence weights are normalised to have an average value of one, the weighted loss \(L_{\text{adj}}\) remains comparable in scale to \(L_{\text{central}}\).  
This joint loss formulation enables the network to integrate fully reliable supervision from the central section with partially reliable pseudo-labelled data from the adjacent sections, thereby improving overall prediction performance while mitigating the adverse effects of inaccurate pseudo labels.

\subsection{Implementation Details}

\label{method:implementation}

\noindent\textbf{Backbone networks.}\quad
To demonstrate the generality of our framework across different network architectures, we implement and test ST-DAI using two different backbone designs: a conventional UNet \cite{ronneberger2015u} and a UNet-based diffusion model \cite{ho2020denoising}. 

To fully leverage the rich histology information available for each section, we also propose a novel feature embedding method. Specifically, for each section, the corresponding histology image is processed by a pre-trained \textit{CONCH} model \cite{lu2024visual} to generate a feature map \(\phi(e) \in \mathbb{R}^{C_e \times \hat{H} \times \hat{W}}\), where \(\hat{H}\) and \(\hat{W}\) represent the spatial dimensions of the histology-derived feature map. This feature map, obtained from the histology image, is then transformed using a convolutional layer to match the spatial dimensions \(H\) and \(W\) of the intermediate feature maps produced by the backbone network. The transformed histology feature map is then concatenated with the network's intermediate feature map \(F \in \mathbb{R}^{C \times H \times W}\) along the channel dimension, resulting in an enhanced feature representation:
\begin{equation}
F' = \operatorname{Concat}\left(F, \phi(e)\right).
\end{equation}
This fusion of histology-derived features with the learned network features enriches the model's ability to capture fine-grained morphological details, significantly improving the accuracy of gene expression prediction.

\noindent\textbf{Training settings.}\quad
All experiments were conducted on a single NVIDIA 4090 GPU using the PyTorch framework. For optimization, we employed the Adam optimizer to update the model parameters. During the Pretraining phase, the UNet model was trained for 500 epochs, while the diffusion model was trained for 1000 epochs. In the FMDR, the UNet model was trained for 1000 epochs, and the diffusion model underwent 2000 epochs. The initial learning rate for all models was set to 0.001, with cosine annealing used to adjust the learning rate dynamically throughout the training process. 

\subsection{Datasets, Baselines, and Evaluation Metrics} \label{method:datasets}

\noindent\textbf{Dataset.}\quad
We evaluate our method on a subset of the HEST\textendash{}1K dataset \cite{Xenium2022}, a diverse collection of ST data.  Specifically, we analyse six high-resolution breast-cancer samples obtained with the Xenium platform, arranged into three pairs of spatially adjacent tissue sections.  The first two pairs, TENX99 with TENX98 and TENX97 with TENX95, originate from two different patients diagnosed with invasive ductal carcinoma, whereas the third pair, TENX96 with TENX94, is derived from a patient with invasive lobular carcinoma.  Because each pair consists of consecutive sections from the same specimen, the data capture intra-sample heterogeneity while maintaining spatial continuity across sections. All experiments are conducted on eight highly clinically relevant genes, namely FOXA1, MDM2, CD68, PGR, MLPH, ESR1, ERBB2, and EGFR.

\noindent\textbf{Baselines.}\quad
For comparative evaluation, we benchmark our method against two baselines, TESLA \cite{hu2023deciphering} and DIST \cite{zhao2023dist}. TESLA employs a variational graph autoencoder that integrates Visium-scale spot expression with histological context and single-cell references to infer sub-spot gene-expression profiles, thereby achieving pixel-level resolution of tumour and micro-environmental cell states.  DIST tackles the sparsity and noise of high-resolution platforms by training a convolutional neural network that jointly exploits local image texture and neighbouring transcriptomic patterns to impute missing counts and denoise observed values, which produces refined spatial maps while preserving tissue boundaries. Both are considered state-of-the-art methods for ST imputations.

\noindent\textbf{Evaluation Metrics.}\quad
To rigorously assess the performance of the proposed method, we employ four widely recognized evaluation metrics: Peak Signal-to-Noise Ratio (PSNR), Structural Similarity Index (SSIM), Mean Absolute Error (MAE), and Pearson Correlation Coefficient (PCC). PSNR quantifies the ratio between the maximum possible power of a signal and the power of distorting noise. SSIM measures the perceptual similarity between the reconstructed and the ground truth images. MAE, defined as the average of the absolute differences between the predicted and actual values, provides a direct measure of overall prediction accuracy. Finally, PCC evaluates the degree of linear correlation between the predicted and ground truth values, offering insight into the consistency between the predictions and the true data. 

\section{Experimental Results}

\subsection{Main Results}

\begin{figure*}[htbp]
    \centering
    \includegraphics[width=1\linewidth]{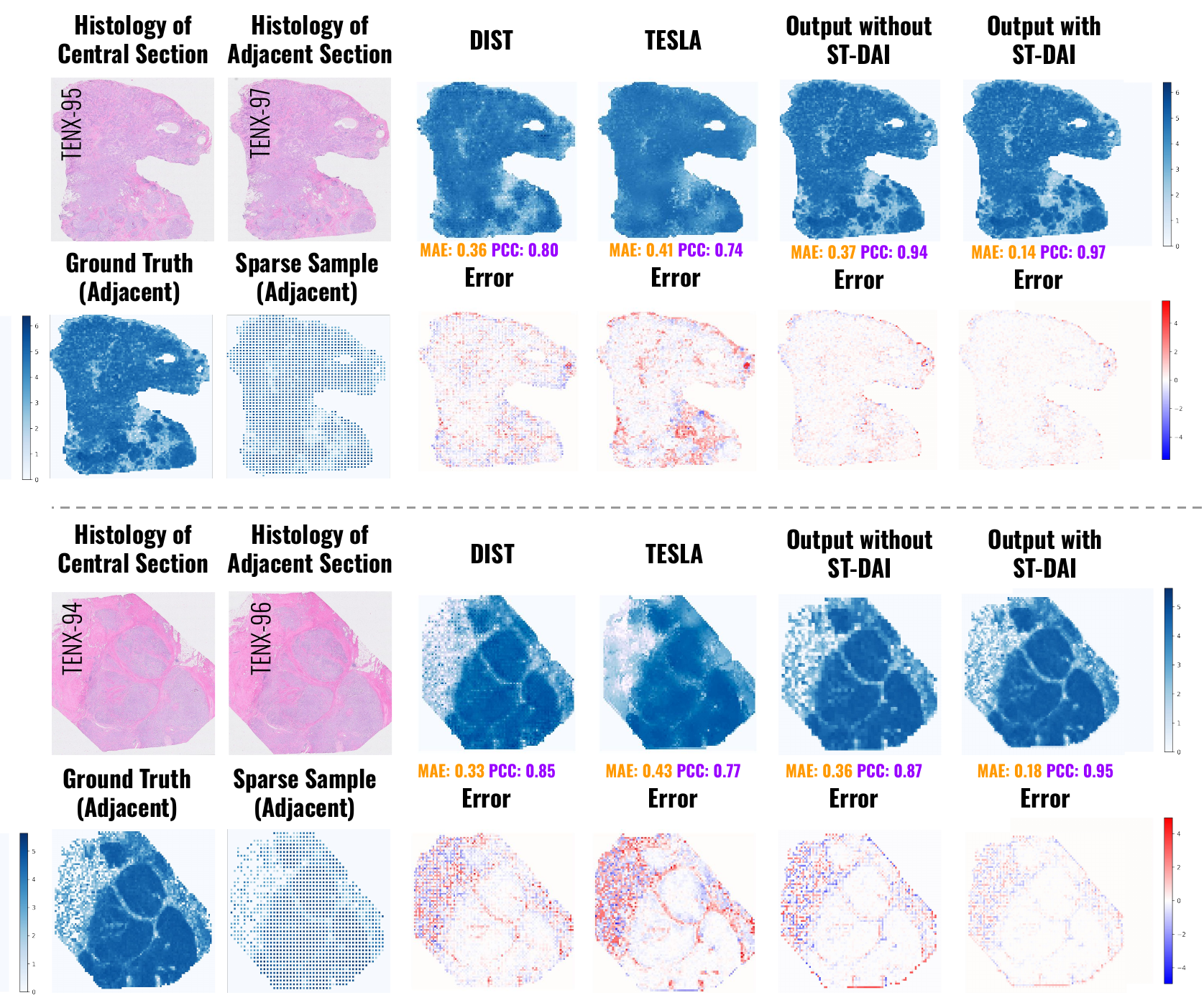}
    \caption{Visual comparison of ERBB2‐expression reconstructions for two adjacent‐section pairs (TENX95–TENX97 and TENX94–TENX96). The figure displays the ground-truth inputs alongside the reconstructions obtained with DIST, TESLA, a vanilla UNet, and the proposed UNet augmented with ST-DAI. An error map annotated with the corresponding MAE and PCC values is shown beneath each reconstruction.}
    \label{fig:vis_comp}
\end{figure*}

Figure~\ref{fig:vis_comp} depicts ERBB2 expression reconstructions for two representative section pairs, TENX95–TENX97 and TENX94–TENX96, from two breast cancer patients.  On the TENX95–TENX97 pair, the TESLA baseline produces visibly blurred maps in which broad tissue regions are either over-saturated or severely attenuated, yielding extensive high-magnitude residuals in the accompanying error plot.  DIST and the vanilla UNet (trained without ST-DAI) offer moderate improvements: the overall texture is sharper and several gross misestimations disappear, yet substantial errors remain, particularly in the peripheral margin regions.  After integrating ST-DAI, the UNet produces markedly sharper spatial detail that aligns closely with the ground-truth expression patterns, and the corresponding residual map shows a pronounced reduction in error magnitude. A similar improvement from our approach is observed for the TENX94–TENX96 pairs.  Collectively, the visual evidence highlights the ability of ST-DAI to mitigate inter-section domain discrepancies and to deliver high-fidelity volumetric reconstructions for gene expression.

\begin{table}[htbp]
\centering
\fontsize{5}{7}\selectfont
\setlength{\tabcolsep}{1pt}
\caption{Quantitative evaluation of gene expression reconstruction performance for UNet and Diffusion networks across various 3D adjacent tissue section pairings. Each entry is presented as $\,\text{mean}\pm\text{SD}\,$, where the mean is computed over all genes and the accompanying value denotes the corresponding standard deviation across those genes. Results are shown for both backbone networks with or without the proposed ST-DAI framework. }
\begin{tabular}{cc | c | cccc | cccc}
\toprule
\multicolumn{2}{c|}{3D section Setting} & Model & \multicolumn{4}{c|}{UNET} & \multicolumn{4}{c}{DIFFUSION} \\
\midrule
Central & Adjacent & ST-DAI & PSNR$\uparrow$ & SSIM$\uparrow$ & MAE$\downarrow$ & PCC$\uparrow$ & PSNR$\uparrow$ & SSIM$\uparrow$ & MAE$\downarrow$ & PCC$\uparrow$ \\
\midrule
\multirow{2}{*}{TENX95} & \multirow{2}{*}{TENX97} &        & $22.61\pm1.26$ & $0.81\pm0.01$ & $0.26\pm0.02$ & $0.95\pm0.01$ & $23.43\pm3.01$ & $0.82\pm0.01$ & $0.24\pm0.02$ & $0.95\pm0.01$ \\
                       &                         & \checkmark & $27.21\pm3.45$ & $0.94\pm0.02$ & $0.13\pm0.01$ & $0.98\pm0.00$ & $28.06\pm2.61$ & $0.95\pm0.01$ & $0.12\pm0.01$ & $0.98\pm0.00$ \\
\midrule
\multirow{2}{*}{TENX97} & \multirow{2}{*}{TENX95} &        & $22.27\pm2.03$ & $0.77\pm0.01$ & $0.28\pm0.02$ & $0.94\pm0.01$ & $22.77\pm2.34$ & $0.82\pm0.02$ & $0.25\pm0.02$ & $0.94\pm0.01$ \\
                       &                         & \checkmark & $27.05\pm2.37$ & $0.94\pm0.01$ & $0.15\pm0.01$ & $0.97\pm0.01$ & $27.60\pm2.46$ & $0.95\pm0.02$ & $0.12\pm0.01$ & $0.98\pm0.00$ \\
\midrule
\multirow{2}{*}{TENX98} & \multirow{2}{*}{TENX99} &        & $17.22\pm1.80$ & $0.70\pm0.02$ & $0.34\pm0.02$ & $0.88\pm0.01$ & $16.03\pm2.41$ & $0.65\pm0.01$ & $0.37\pm0.01$ & $0.87\pm0.01$ \\
                       &                         & \checkmark & $24.73\pm2.06$ & $0.91\pm0.01$ & $0.17\pm0.01$ & $0.96\pm0.01$ & $22.66\pm1.65$ & $0.86\pm0.02$ & $0.20\pm0.02$ & $0.95\pm0.00$ \\
\midrule
\multirow{2}{*}{TENX99} & \multirow{2}{*}{TENX98} &        & $17.13\pm1.47$ & $0.70\pm0.01$ & $0.33\pm0.01$ & $0.89\pm0.01$ & $15.49\pm2.62$ & $0.60\pm0.02$ & $0.39\pm0.02$ & $0.87\pm0.01$ \\
                       &                         & \checkmark & $25.67\pm2.40$ & $0.90\pm0.01$ & $0.16\pm0.01$ & $0.96\pm0.01$ & $21.63\pm3.06$ & $0.83\pm0.01$ & $0.23\pm0.02$ & $0.93\pm0.01$ \\
\midrule
\multirow{2}{*}{TENX94} & \multirow{2}{*}{TENX96} &        & $16.05\pm1.43$ & $0.63\pm0.02$ & $0.37\pm0.01$ & $0.87\pm0.01$ & $16.12\pm2.07$ & $0.61\pm0.01$ & $0.38\pm0.02$ & $0.87\pm0.00$ \\
                       &                         & \checkmark & $23.75\pm2.51$ & $0.89\pm0.03$ & $0.18\pm0.01$ & $0.96\pm0.01$ & $22.24\pm1.48$ & $0.86\pm0.02$ & $0.21\pm0.01$ & $0.95\pm0.01$ \\
\midrule
\multirow{2}{*}{TENX96} & \multirow{2}{*}{TENX94} &        & $16.37\pm2.11$ & $0.65\pm0.01$ & $0.36\pm0.02$ & $0.86\pm0.01$ & $15.20\pm1.06$ & $0.58\pm0.01$ & $0.40\pm0.01$ & $0.83\pm0.01$ \\
                       &                         & \checkmark & $24.61\pm1.44$ & $0.90\pm0.02$ & $0.17\pm0.01$ & $0.96\pm0.01$ & $20.34\pm2.43$ & $0.80\pm0.01$ & $0.25\pm0.02$ & $0.94\pm0.00$ \\
\bottomrule
\end{tabular}
\label{tab:exp_results}
\end{table}

Table~\ref{tab:exp_results} summarises the mean reconstruction performance for all profiled genes across every section–pair configuration, comparing a baseline model with its ST-DAI–enhanced counterpart as indicated by the absence or presence of a check-mark.  Averaged over all genes, the inclusion of ST-DAI consistently yields higher PSNR and SSIM together with lower MAE and higher PCC for every section pairing.  On the TENX95–TENX97 pair, for example, UNet performance improves from $22.61$ to $27.21$\,dB in PSNR and from $0.81$ to $0.94$ in SSIM, while MAE drops from $0.26$ to $0.13$ and PCC rises from $0.95$ to $0.98$.  With the Diffusion backbone the same pair shows gains from $23.43$ to $28.06$\,dB in PSNR, from $0.82$ to $0.95$ in SSIM, and corresponding reductions in MAE and increases in PCC.  Similar improvements appear in all other configurations, such as TENX98–TENX99 where UNet PSNR rises from $17.22$ to $24.73$\,dB, SSIM from $0.70$ to $0.91$, MAE falls from $0.34$ to $0.17$, and PCC climbs from $0.88$ to $0.96$.  Comparable trends are observed for the Diffusion model and for the TENX94–TENX96 and TENX96–TENX94 pairs.  The uniform enhancement in every metric demonstrates that ST-DAI systematically improves reconstruction fidelity and mitigates errors, thereby overcoming the limitations of conventional sampling and inference strategies. 

\begin{table}[htbp]
\centering
\caption{Performance comparison between the ST-DAI–enhanced UNet and two alternative reconstruction methods across multiple adjacent-section configurations.}
\label{tab:method_comparison}
\begin{tabular}{cc|c|cccc}
\toprule
\multicolumn{2}{c|}{\textbf{3D section Setting}} & \multirow{2.5}{*}{Method} & \multirow{2.5}{*}{PSNR$\uparrow$} & \multirow{2.5}{*}{SSIM$\uparrow$} & \multirow{2.5}{*}{MAE$\downarrow$} & \multirow{2.5}{*}{PCC$\uparrow$} \\
\cmidrule{1-2}
Central & Adjacent & & & & & \\
\midrule
\multirow{3}{*}{TENX95} & \multirow{3}{*}{TENX97} & TESLA        & 16.88 & 0.56 & 0.41 & 0.74 \\
                        &                          & DIST         & 17.12 & 0.63 & 0.36 & 0.80 \\
                        &                          & \textbf{Ours}& 28.04 & 0.94 & 0.14 & 0.97 \\
\midrule
\multirow{3}{*}{TENX97} & \multirow{3}{*}{TENX95} & TESLA        & 13.75 & 0.52 & 0.45 & 0.71 \\
                        &                          & DIST         & 16.65 & 0.60 & 0.37 & 0.80 \\
                        &                          & \textbf{Ours}& 25.73 & 0.94 & 0.16 & 0.97 \\
\midrule
\multirow{3}{*}{TENX94} & \multirow{3}{*}{TENX96} & TESLA        & 16.04 & 0.51 & 0.43 & 0.77 \\
                        &                          & DIST         & 17.31 & 0.67 & 0.33 & 0.85 \\
                        &                          & \textbf{Ours}& 24.20 & 0.88 & 0.18 & 0.95 \\
\midrule
\multirow{3}{*}{TENX96} & \multirow{3}{*}{TENX94} & TESLA        & 16.22 & 0.52 & 0.40 & 0.75 \\
                        &                          & DIST         & 18.09 & 0.69 & 0.31 & 0.86 \\
                        &                          & \textbf{Ours}& 24.49 & 0.91 & 0.17 & 0.95 \\
\bottomrule
\end{tabular}
\end{table}

Table~\ref{tab:method_comparison} presents the reconstruction performance for the ERBB2 gene across four adjacent-section settings, comparing TESLA\,\cite{hu2023deciphering}, DIST\,\cite{zhao2023dist}, and the proposed UNet trained with the ST-DAI domain-adaptive framework. The ST-DAI model employs the same UNet architecture as the baselines, whose improvements arise solely from the incorporation of section-specific alignment and parameter-efficient calibration layers introduced by ST-DAI. On the TENX95 to TENX97 pair, TESLA achieves a PSNR of 16.88\,dB, an SSIM of 0.56, an MAE of 0.41, and a PCC of 0.74, while DIST attains 17.12\,dB, 0.63, 0.36, and 0.80.  The ST-DAI-augmented UNet increases the PSNR to 28.04\,dB, elevates the SSIM to 0.94, reduces the MAE to 0.14, and lifts the PCC to 0.97.  Similar margins are observed for the remaining section pairs.  These results demonstrate that ST-DAI markedly enhances reconstruction fidelity by effectively aligning and adapting feature representations across adjacent tissue domains.

\subsection{Ablation Study}
All ablation experiments were conducted under a unified setting using TENX95 as the central section and TENX97 as the adjacent section with a UNet backbone on all genes. The training configurations, including hyperparameters and convergence criteria, were kept identical across all experiments to ensure fairness, and each model was trained until complete convergence was achieved. 

\begin{table*}[htbp]
\centering
\setlength{\tabcolsep}{2pt}
\caption{Ablation Studies for TENX95 (Central section) and TENX97 (Adjacent section) with UNet. \textsuperscript{\dag}FMDR denotes only the dual-branch joint fine-tuning method (excluding PDL and CSG). }
\label{tab:ablation}
\begin{tabular}{c|ccc|c|ccccc}
\toprule
CSA & FMDR\textsuperscript{\dag} & PDL & CSG & DCO & PSNR$\uparrow$ & SSIM$\uparrow$ & MAE$\downarrow$ & PCC$\uparrow$ \\
\midrule
\xmark & \xmark & \xmark & \xmark & \xmark & $22.61\pm1.26$ & $0.81\pm0.01$ & $0.26\pm0.02$ & $0.95\pm0.01$ \\
\xmark & \checkmark & \xmark & \xmark & \xmark & $20.35\pm2.02$ & $0.75\pm0.02$ & $0.28\pm0.02$ & $0.91\pm0.01$ \\
\checkmark & \xmark & \xmark & \xmark & \xmark & $23.01\pm1.39$ & $0.81\pm0.02$ & $0.24\pm0.01$ & $0.93\pm0.01$ \\
\checkmark & \checkmark & \xmark & \xmark & \xmark & $23.00\pm1.90$ & $0.81\pm0.01$ & $0.25\pm0.02$ & $0.93\pm0.01$ \\
\checkmark & \checkmark & \checkmark & \xmark & \xmark & $23.72\pm2.55$ & $0.87\pm0.02$ & $0.20\pm0.01$ & $0.84\pm0.01$ \\
\checkmark & \checkmark & \checkmark & \checkmark & \xmark & $24.76\pm2.24$ & $0.91\pm0.01$ & $0.17\pm0.01$ & $0.96\pm0.00$ \\
\checkmark & \checkmark & \checkmark & \checkmark & \checkmark & $27.21\pm3.45$ & $0.94\pm0.02$ & $0.13\pm0.01$ & $0.98\pm0.00$ \\
\bottomrule
\end{tabular}
\end{table*}

Table~\ref{tab:ablation} presents the results of the ablation study conducted on the TENX95 (central section) and TENX97 (adjacent section) pair using a UNet backbone. The table evaluates the impact of various components on gene expression prediction performance, including Cross-section Alignment (CSA), FMDR\textsuperscript{\dag} (only the dual-branch joint fine-tuning method, excluding PDLs and CSG), PDL, CSG, and Data Consistency Operation (DCO). 

The first row of the table shows the baseline performance when none of the modules are enabled. In this scenario, the model achieves a PSNR of 22.61 dB, an SSIM of 0.81, an MAE of 0.26, and a PCC of 0.95. These results represent the performance of the model when no domain-specific adjustments or enhancements are applied. When Cross-section Alignment (CSA) is introduced, as seen in the third row, the PSNR improves slightly to 23.01 dB and SSIM remains 0.81. This indicates that aligning the central and adjacent sections does have a positive impact on performance by ensuring that the feature representations of the sections are consistent. Although this is a modest improvement, it demonstrates the effectiveness of CSA in reducing misalignments between sections.

However, when FMDR is introduced in row 2 without the inclusion of PDL and CSG, the performance deteriorates. Specifically, PSNR decreases from 22.61 to 20.35, SSIM drops from 0.81 to 0.75, and MAE increases from 0.26 to 0.28. This indicates that without appropriate domain alignment (via PDL) and pseudo-label reliability assessment (via CSG), multi-domain joint fine-tuning not only fails to improve performance, but also potentially exacerbates the issue. The misalignment between the central and adjacent sections results in suboptimal learning and inaccurate gene expression predictions. A similar trend is observed in row 4, where CSA is included, but FMDR is applied without PDL or CSG. In this configuration, despite the alignment of the sections, the performance still decreases, with PSNR remaining at 23.00, SSIM at 0.81, and MAE at 0.25. This further emphasizes the crucial role of domain alignment and pseudo-label confidence in optimizing multi-domain fine-tuning, showing that without these essential components, the potential benefits of multi-domain joint training are not realized, and performance can actually suffer.

In contrast, when PDL is incorporated in row 5, the performance improves significantly. The PSNR increases to 23.72 dB, and SSIM increases to 0.87, demonstrating the effectiveness of PDL in recalibrating the shared feature representations between sections. PDL ensures that the features from the central section are adapted to the characteristics of the adjacent section without disturbing the pretrained network’s feature distribution. Further, the introduction of CSG in row 6 leads to additional improvements, with PSNR rising to 24.76 dB and SSIM reaching 0.91. The CSG module provides a crucial role in weighting the importance of different regions of the adjacent section, which helps the model focus on the more reliable areas during training and further enhances the accuracy of the predictions.

Finally, the complete configuration in row 7, where all modules are activated, achieves the highest performance with a PSNR of 27.21 dB, an SSIM of 0.94, an MAE of 0.13, and a PCC of 0.98. This demonstrates that the combined effects of CSA, FMDR, PDL, CSG and DCO result in a substantial enhancement in performance, confirming the complementary benefits of each component within our framework. Moreover, the integration of PDL and CSG is crucial for effective domain adaptation, ensuring accurate multi-domain joint fine-tuning. While FMDR alone offers measurable benefits, its combination with PDL and CSG further enhances the model’s capacity to mitigate domain discrepancies and, in turn, improves gene-expression prediction accuracy. The findings confirm that the combined approach of FMDR, PDL, and CSG leads to optimal performance in gene expression reconstruction.

\section{Discussion}
In summary, the experimental evidence presented in Figure~\ref{fig:vis_comp} and Tables~\ref{tab:exp_results}--\ref{tab:ablation} demonstrates that ST-DAI constitutes a potential practical and scalable solution for volumetric ST. The framework combines a cost-efficient 2.5D selective sampling strategy with a intra-sample domain-adaptive 3D imputation pipeline.  The sampling component acquires complete measurements on the central section and collects sparse observations on the adjacent sections, thereby markedly reducing sequencing cost.  The imputation component then mitigates intra-sample domain gaps by aligning sections, generating confidence-weighted pseudo supervision, and refining only a few parameters through Fast Multi-Domain Refinement.  Quantitative evaluations show consistent improvements in PSNR, SSIM, MAE, and PCC across all section pairs, and qualitative inspection shows that the method recovers fine-grained gene expression patterns that baseline approaches fail to resolve.  Ablation studies further verify that Cross-section Alignment (CSA), Parameter-Efficient Domain-Alignment Layers (PDL), Confidence Score Generator (CSG), and Data Consistency Operation (DCO) each contribute to performance gains.  Our method demonstrates a practical and scalable approach for volumetric ST profiling, offering significant improvements in gene expression prediction with reduced experimental costs. While ST-DAI demonstrates clear benefits, some limitations remain that warrant further investigation. 

First, the current Cross-section Alignment (CSA) method requires the central section to be aligned with each adjacent section individually. This approach necessitates multiple training steps when more than one adjacent section is involved, which could be computationally inefficient. An alternative approach could be to align all adjacent sections to the central section simultaneously. However, this solution may introduce inconsistencies between the training and inference processes, as the adjacent sections used for inference would not undergo the same alignment transformation as during training. An alternative potential solution could involve adding an additional branch in FMDR, where images are randomly rotated by small angles (simulating the alignment transformation) before being input into the model. By employing adversarial learning \cite{park2024adversarial, li2018domain, zhang2021joint, zhang2020adversarial, wang2020adversarial, tzeng2017adversarial, jang2022unknown, qian2025ama} or contrastive learning methods \cite{quintana2025bridging, liu2021margin, en2024unsupervised, gu2022contrastive, thota2021contrastive, tang2023self, shen2022explaining}, this branch could potentially align the feature representations of the same section before and after rotation, ensuring consistency across sections and reducing the need for multiple training iterations, which will be an important future direction. 

Second, the current sampling of adjacent sections relies on a fixed strategy that merely doubles the sampling interval. Although this reduces sampling cost, it ignores the spatial variability in prediction difficulty. Complex regions, such as highly heterogeneous areas, pose greater challenges, whereas homogeneous regions are comparatively easy to predict.  A more refined, adaptive sampling scheme could allocate additional measurements to regions where larger errors are expected and fewer to regions that are readily predictable, thereby improving overall accuracy while avoiding unnecessary computational cost.  Moreover, complementary enhancements include region-specific attention mechanisms that direct the network to focus on clinically critical areas during training \cite{lee2020structure, chen2021medical}, as well as targeted loss functions that impose higher penalties on errors within difficult regions \cite{eljurdi2021surprisingly, hatamizadeh2019end}.  Together, these strategies potentially promise to balance predictive performance and resource consumption on large-scale datasets, may deliver more accurate and clinically meaningful results, and may ultimately increase the utility of ST models in real-world medical settings.

\section{Conclusion}
This study introduces ST-DAI, a unified framework that couples a cost-efficient 2.5D sampling protocol with a intra-sample domain-adaptive imputation pipeline to enable accurate 3D reconstruction of ST from minimally sampled data. The proposed sampling scheme exhaustively sequences the central section and sparsely samples adjacent sections, thereby substantially lowering sequencing requirements. Building on the customized 2.5D acquisition scheme, the proposed imputation pipeline aligns sections, generates confidence-aware pseudo supervision, and performs parameter-efficient refinement, collectively reducing inter-section domain gaps and yielding accurate gene-expression imputation. Comprehensive experiments demonstrate that the proposed framework consistently attains superior quantitative performance than both the corresponding backbone trained without this framework and other baseline approaches, while accurately recovering fine-grained gene-expression patterns across diverse samples. By showing that accurate 3D gene-expression landscapes can be reconstructed from cost-effective acquisitions, ST-DAI establishes a promising practical foundation that is potentially useful for routine large-scale ST investigations and broadens the scope of 3D ST analysis in both fundamental research and clinical practice.

\bibliography{sn-bibliography}

\end{document}